\documentclass[12pt]{amsart}

\numberwithin{equation}{section}

\newcommand{\bE}{\mathbb{E}}

\newcommand{\bR}{\mathbb{R}}

\newcommand{\cF}{\mathcal{F}}
\newcommand{\cG}{\mathcal{G}}

\theoremstyle{plain} 
\theoremstyle{plain} 
\theoremstyle{plain} 
\theoremstyle{plain} 
\theoremstyle{remark} 
\theoremstyle{definition} 
\theoremstyle{definition} 
\theoremstyle{definition} 
\theoremstyle{remark}

\begin{document}

\title[Divergence Times]{Unidentifiable divergence times in rates--across--sites models}

\author{Steven N.\ Evans}
\address{Department of Statistics \#3860 \\
  University of California at Berkeley \\
367 Evans Hall \\
Berkeley, CA 94720-3860 \\
U.S.A}
\email{evans@stat.Berkeley.EDU, phone: 1-510-642-2777, fax: 1-510-642-7892}
\thanks{SNE is the Corresponding Author and was supported in part by NSF grants DMS-0071468 and DMS-0405778}

\author{Tandy Warnow}
\address{Department of Computer Sciences \\
University of Texas at Austin \\
Austin, TX 78712 \\
U.S.A.}
\email{tandy@cs.utexas.edu, phone: 1-512-471-9724, fax: 1-512-471-8885}
\thanks{TW was supported by NSF grants EF-0331453, BCS-0312830, and IIS-0121680}

%\date{\today}


\begin{abstract}
The {\em rates--across--sites} assumption in 
phylogenetic inference posits that
the rate matrix governing the Markovian evolution
of a character on an edge of the putative phylogenetic tree
is the product of a character-specific scale factor
and a rate matrix that is particular to
that edge.
Thus,  evolution follows basically the
same  process for all characters,
except that it occurs faster for some characters
than others.  
To allow estimation of tree topologies and edge lengths for
such models, it is commonly assumed that the scale factors
are not arbitrary unknown constants,  but rather
unobserved, independent, identically distributed draws
from a member of some parametric family of distributions.  A popular
choice is the gamma family.  We consider an example of a clock-like tree with
three taxa, one unknown edge length, a known root state, and a parametric family
of scale factor distributions that contain the gamma family.  
This model has the property that, for a generic choice of unknown
edge length and scale factor distribution, there is another edge length
and scale factor distribution which generates data with exactly
the same distribution, so that even with infinitely many data it
will be typically impossible to make correct inferences about the unknown
edge length.
\end{abstract}

\maketitle

\baselineskip=21pt

\section{Introduction}
Beginning with the germinal work \cite{Fel78},
statistically-based
estimations of phylogenetic trees have become
popular in molecular systematics, with
Bayesian  \cite{HuelsenbeckRonquist2001}
and maximum likelihood methods \cite{Swofford96,GuindonOlivier2003,KosakovskyPondMuse2000,Lewis98,OlsenMHO94}
used with increasing  frequency.
Such statistically-based methods assume that the 
observed sequences are the result of a stochastic process that
has operated on a tree, and they make assumptions about the
stochastic process (that is, model) that has produced the data.  

A fundamental question about any statistical model is whether it is
identifiable: that is, whether different parameter values lead to
different probability distributions for the data, so that, in particular, there
is some hope of estimating the parameters with increasing accuracy
as the amount of data increases.
These questions have been investigated extensively for
certain models used in phylogenetic inference (see, for example,
\cite{Ste94, Cha96}).

Many models used in phylogenetic inference combine a (typically
parameter rich) model of individual site evolution with the assumption
that the different sites evolve under a {\em rates--across--sites} model, so that
 each site $c$ has an associated rate of evolution $r_c$ which
is constant across the tree.  Thus, sites evolve under essentially the same 
evolutionary process, and are just scaled up (or down) versions of each other.
(Thus, the rates--across--sites assumption implies that if one site is 
expected to evolve twice as fast
as another site on edge $e$, then it 
is expected to evolve twice as fast as the other site on every edge.)

Such rates--across--sites models in which each
character has its own unknown scale factor
are discussed in \cite{SwoOlsWadHil96}, but these
models still pose difficult inferential problems.
As remarked in Chapter 13 of \cite{Fel04}:
\begin{quote}
As the number of sites increases, the number of parameters being estimated 
rises correspondingly.  This is worrisome: in such
``infinitely many parameters'' cases maximum likelihood
often misbehaves and fails to converge to the correct tree as the
number of sites increases.
\end{quote}
Indeed, our own example below shows
that relative edge-lengths are, in general,
unidentifiable for such models.  (We discuss a situation
in which the unknown scale parameters for the respective characters
are unobserved, independent, identically distributed,
realizations of some  distribution belonging to a
particular family of distributions.  However, if edge-lengths are
not identifiable in our set-up, then they certainly won't be
identifiable in the analogous set-up where the scale parameters
are arbitrary.)

A popular
`fix' that has been proposed for this problem
is to adopt a {\em random effects} approach and
suppose that the successive scale factors
$r_c$ are unobserved, independent random draws from
a member of some parametric family of distributions.  This
reduces the dimensionality of the problem by replacing
the deterministic sequence of $r_c$ parameters with the small
number of parameters that describe the generating distribution
(see, for example, \cite{UzzCor71, NeiChaFue76, Ols87, HasKisYan87,
Yan93}).

A standard choice of distribution for the random scale
factors is the two-parameter family of gamma distributions.  This
family has the mathematical advantage that likelihoods
still have analytically tractable closed forms, and  it was shown
for a wide class of substitution models in \cite{Rog01} that edge-lengths
are identifiable in this setting.
The choice of the gamma family is
often supported by claims that it is sufficiently flexible
to mimic the variation of rates between characters
that is likely to be seen
`in practice'.  
There also appears to be a general
sense among many practitioners that the choice of distributional
family
for the scale factors is primarily a matter of convenience and that,
provided the family is rich enough, substantially
correct inferences of relative edge-lengths will
be possible with sufficient data.  To our knowledge,
there is no argument in any setting justifying why an assumption
of an exact gamma distribution for the scale factors
is biologically reasonable.  As remarked in \cite{Fel04}:
\begin{quote}
There is nothing about the gamma distribution that makes it more
biologically realistic than any other distribution, such as the
lognormal.  It is used because of its mathematical tractability.
\end{quote}

It was shown in \cite{SteSzeHen94} that the use of
random scale factors might not be completely without problems.  
In their paper they gave an example 
  of a specific choice of edge-lengths for each
tree topology and a specific choice of discrete distribution
for the scale factors (rather than a continuous
distribution such as a gamma) such that the resulting
distribution for the data under the Neyman
two-state model is the same for all tree topologies.
%This result means that under the discrete distribution model,
%model trees   are not identifiable.  However, the result does not
%show that   {\em all} model trees (i.e., all  trees with
%particular discrete distribution for the scale factors)
%are not identifiable.

In this paper we go further, at least
in some directions.  We consider
%\begin{itemize}
%\item
a rooted tree with three taxa and one unknown edge-length
(with the remaining edge-lengths either known or
fixed by the clock-like constraint that all lineages have the same
total length),
%\item
and a particular ten-parameter family $\cF$ of scale factor distributions with
a certain nine-parameter sub-family $\cG$ of $\cF$.
%\end{itemize}
We show that for a {\em generic} choice of unknown
edge-length $\tau$ and model
in $G \in \cG$ there is a choice of edge-length $\sigma \ne \tau$
and model in $F \in \cF$ with the property that
%\begin{itemize}
%\item
data generated according to the Neyman 2-state model 
with known ancestral state,
scale factor distributed according to $G$, and edge-length $\tau$,
has the same distribution as 
data generated according to a Neyman 2-state model 
with the same ancestral state, scale factor distributed 
according to $F$, and edge-length $\sigma$. 
Thus,
even with infinitely many data, it would be
impossible to decide whether the unknown edge-length 
is $\sigma$ or $\tau$ -- even if one somehow knew in advance
that the distribution for the scale parameter was one
of either $F$ or $G$.
%\end{itemize}
Here the term {\em generic} means that the set of exceptional
models  $G$ and edge-lengths $\tau$ for which no corresponding 
$F$ and $\sigma$ exist is a lower dimensional subset of
$\cG \times \bR_+$.  In
particular, the set of $G$ and $\tau$ that have corresponding
$F$ and $\sigma$ is an everywhere dense open subset
of $\cG \times \bR_+$.

Moreover, the family $\cF$  consists of distributions with smooth,
unimodal, densities that possess moments of all
orders.  In this sense, each distribution in
$\cF$ is as ``nice'' as a gamma distribution.  Of course, the
two-parameter family of gamma distributions is simpler than
than the ten-parameter family $\cF$.  However, the use of 
$\cF$ is essentially a technical device in our analysis.  
We could have described our results by simply saying that for
a generic unknown edge-length $\tau$ there is a corresponding
edge-length $\sigma \ne \tau$, and two scale parameter distributions
$G$ and $F$, such that
data generated according to the Neyman 2-state model 
with known ancestral state,
scale factor distributed according to $G$, and edge-length $\tau$,
has the same distribution as 
data generated according to a Neyman 2-state model 
with the same ancestral state, scale factor distributed 
according to $F$, and edge-length $\sigma$.  In particular,
unidentifiability is not inherently a case of ``over-parametrization'':
the effect can be produced when we have just a finite number of
possible parameter values and is not produced by having a continuous
space of possible parameter values with too high a dimension.  We have
included the mention of the families $\cF$ and $\cG$ in the description
of our results to stress that the unidentifiability problem is, in some
sense, generic.

The family $\cG$ (and hence $\cF$) contains all the gamma distributions as
a subfamily.  Any gamma distribution and any
edge-length $t$ will have 
a distribution $G \in \cG$ and $\tau$
arbitrarily close to them such that there is
a corresponding $F \in \cF$ and  $\sigma \ne \tau$
as above.  

Our example applies not only to the
Neyman model but also to any model such
as the binary General Time-Reversible model that contains
the Neyman model as a sub-model.  Furthermore,
our example applies to the General Time-Reversible
on an arbitrary finite state-space, because one can choose
the substitution rate matrices for such a model to be sufficiently
symmetric so that a suitable many-to-one binary encoding of
the model is Markovian and evolves according to the Neyman model.

We should point out that we construct our example using
a perturbative technique.  Consequently, the edge-lengths
$\tau$ and $\sigma$ that arise will be ``close''
to each other.  However, our analysis doesn't rule out
the possibility that a similar example could be
produced with edge-lengths that are ``far apart''.  
In order to fully assess the practical implications of
the phenomenon we have observed, further research is necessary
to quantify just how distant two edge-lengths can be
and still have corresponding scale parameter distributions
that lead to identical distributions for the data.  Moreover,
this is not a purely mathematical question, because the notions
of ``close'' and ``far apart'' are dependent on the 
scientific question being investigated with a particular data set.

Also, we note that if we actually knew the distribution of the 
scale parameter in our three taxa example, then the unknown
edge-length could be recovered uniquely from the distribution
of the data, and this is so for an arbitrary scale parameter
distribution, not just the ones we consider in this paper.
Moreover, the functional that recovers the unknown edge-length is
continuous in the scale parameter distribution when
one equips the space of distributions with the usual topology of
weak convergence.  This suggests that if
we somehow knew the scale parameter distribution up to
some small error, then this would constrain the errors
we could make in determining the edge-length.  However, it
is not clear how well one can identify the relevant features of the scale parameter
distribution: the functional that recovers
the unknown edge-length depends on the functional inverse
of the Laplace transform of the scale parameter distribution
and hence, {\em a priori}, on the entirety of the distribution rather than
some finite dimensional set of features such as the first few moments,
and so there is an apparent need to estimate the whole distribution
quite well.  Once again, this is not solely a theoretical matter
and the extent to which this continuity observation is relevant will depend  
partly on context.

%In conclusion, the attempt to achieve
%identifiability and reasonable inference of
%edge-lengths in the rates--across--sites model
%by using random scale factors that come from some
%common distribution can be problematic.

The rest of the paper is organized as follows. We begin with
an introduction of the mathematical terms in Section \ref{section:basics},
and we present our
example  in Section \ref{section:example}.  
We conclude with a
discussion of the ramifications of
this result, and 
directions for future research in Section \ref{section:conclusion}.

\section{Basics}
\label{section:basics}
In phylogenetic inference, the data are the respective states
of an ensemble of characters exhibited by each of a collection
of taxa.  The most commonly used statistical models in the area
are parameterized by a rooted tree with edge-lengths (which typically represent
the expected number of times a site changes
on the edge when the substitution mechanism is in equilibrium) and a set of Markovian stochastic
mechanisms for the evolution of successive characters down the tree.
It is usually assumed that the observed states for different
characters are statistically independent.  The goal of phylogenetic
inference is to estimate some or all of: the shape (topology) of the
tree, the lengths  of the edges, and 
any unknown parameters involved in the specification of the
evolution mechanism.

We will restrict attention to the case where each character has
the same finite set of possible states. For example, the characters
could be nucleotides exhibited at different sites on
the genome, and so each character is in one of the four
states $\{A,G,C,T\}$.  In the example we will give in this paper,
we will work with (binary) characters having one of two possible
states, $0$ or $1$.
 For each character $c$ and each
edge $e$ in the tree, one then has a rate matrix $Q_{c,e}$
that describes the evolutionary process on edge $e$ for
character $c$ (we refer the reader unfamiliar with continuous
time Markov chains to a standard text such as
\cite{GriSti01}).  Thus, 
given that the
character is in state $i$ at the beginning of
the edge, the conditional probability 
of the (possibly unobserved) event
that  it is in state $j$ at the end of the edge
is the $(i,j)$ entry of the
matrix exponential $\exp(t Q_{e,c})$, where
$t$ is the length of $e$.  The matrix $Q_{c,e}$
has row sums equal to $0$ and non-negative off-diagonal
entries: $-Q_{c,e}(i,i)$ is the rate at which the character leaves
the state $i$ and $-Q_{c,e}(i,j)/Q_{c,e}(i,i)$ is the probability
that it jumps to state $j$ when it leaves state $i$.

Single site substitution
models can range from the very
simple (e.g.  the Jukes-Cantor and Kimura 2-parameter models) to 
the very complex (e.g. the General Markov Model), which,  for a fixed
character $c$, allow
the $Q_{e,c}$ matrices to vary significantly from
edge to edge, and to have many free parameters.
However, the variation between the different matrices 
obtained by varying the character $c$ is typically more
proscribed.
The most complex model is
where there are no constraints placed on the $Q_{c,e}$; this is
called the ``no common mechanism model" \cite{tuffley.steel}.  Under
this no common mechanism model, it 
will clearly be difficult to recover any information
about edge-lengths.  A simple class of models in
which it is possible to extract information about
edge-lengths is the class in which $Q_{c,e}$ is
the same for all characters $c$ and edges $e$.  Even
for this simple model, there is -- as is well-known --
a certain lack of identifiability, because the same
probability distribution for the data would arise if the
common rate matrix was  multiplied by a common
scale factor and all edge-lengths were divided
by that same factor.  Thus, even for this model
one can only hope to make inferences about relative
edge-lengths unless at least one edge-length is
assumed to be known.  

The more commonly used models assume that the
different $Q_{c,e}$ matrices are themselves the
product of a rate matrix specific to the edge
$e$, and a scale factor that is specific to the character
$c$.  Thus, the evolutionary process that governs one
character is identical, up to a scalar multiple, to that governing 
another character.  
This is the {\em rates--across--sites} assumption in molecular 
phylogenetics, and it has the rather strong implication that
if a character $c$ is expected to evolve twice as fast on edge $e$
as character $c'$, then $c$ is expected to evolve twice as fast on
{\em every} edge in the tree.

The assumption of a common
rate matrix for all  edges is the
{\em molecular clock} assumption, which is
known to be untenable in many situations \cite{JinNei90, Ree92}.
Perhaps the next simplest class of models is the family of
{\em rates-across-sites} models in which $Q_{c,e}$
is the product of a character-specific scale factor
and a rate matrix that is common to all characters
and edges.  That is, $Q_{c,e}$ is of the form $r_c \bar Q$.
In other words, evolution follows basically the
same pattern on all lineages for all characters,
except that it occurs faster for some characters
than others.  

Because of the  inferential difficulties of allowing the rates for the different
sites to be arbitrary, these random scale factors are typically assumed to be
drawn from a distribution. Of the many possible distributions, the
most popular distributions are
the  two-parameter   gamma distributions. 
In fact, in practice, almost all estimations of phylogenetic trees are
based upon the assumption that the rates across sites are drawn from
a gamma distribution, or a discretized  gamma
distribution.  Also, it is sometimes assumed that certain characters
are invariable (that is, that the scale parameter for such sites is $0$).

\section{The example}
\label{section:example}

We will present an example of a tree with three
taxa, and with sites evolving under the
Neyman two-state model (i.e., the two-state version of the Jukes-Cantor
model of evolution) with a known state at the root.

Consider a tree with three taxa $x$, $y$, and $z$,
a root $v$, and internal node $w$ that is ancestral
to $x$ and $y$.  
The edges $(w,x)$ and $(w,y)$
have a known length, which we can take
as $1$.  Suppose further that the edge
$(v,w)$ has unknown length $\sigma$
and that the the tree is clock-like,
so that the edge $(v,z)$ has length $\sigma+1$.

Suppose there are $\{0,1\}$-valued
characters labelled $1,2,\ldots$
that have evolved on this tree. The
$i^{\mathrm{th}}$ character evolves
according to the Neyman model
with rate $r_i$.  That is, the transition
matrix for an edge of length $t$ is
\[
\frac{1}{2}
\begin{pmatrix}
(1 + \exp(-2 r_i t)) & (1 - \exp(-2 r_i t))\\
(1 - \exp(-2 r_i t)) & (1 + \exp(-2 r_i t)) \\
\end{pmatrix}
=:
\begin{pmatrix}
p_t^{(i)}(0,0) & p_t^{(i)}(0,1) \\
p_t^{(i)}(1,0) & p_t^{(i)}(1,1)
\end{pmatrix},
\]
say.

The probability distribution for the $i^{\mathrm{th}}$
character (that is, the marginal likelihood for this
character) is given as follows.  Suppose it is known that the
state $s_v \in \{0,1\}$ is exhibited by the root $v$. Then the probability
that states $s_x$, $s_y$, and $s_z$ are exhibited by the
taxa $x$, $y$, and $z$ is
\[
\sum_{s_w \in \{0,1\}} p_\sigma^{(i)}(s_v, s_w) p_1^{(i)}(s_w, s_x) p_1^{(i)}(s_w, s_y)
p_{\sigma+1}^{(i)}(s_v, s_z).
\]
Assume that successive characters evolve independently.

The probability distribution for the $i^{\mathrm{th}}$
character is thus easily seen to be a linear combination of the terms
\[
\begin{matrix}
 & 1 & \\
\exp(-2 r_i) & \exp(-4 r_i) & \exp(-2 r_i \sigma) \\
\exp(-2 r_i (1+\sigma)) & \exp(-2 r_i (2+\sigma)) &\exp(-2 r_i (3+\sigma)) \\
\exp(-2 r_i (1+2\sigma)) & \exp(-2 r_i (2+2\sigma)) & \exp(-2 r_i (3+2\sigma))
\end{matrix}
\]
As one of the referees of this paper remarked, by explicitly writing out the
likelihood or using Corollary 8.6.6 of \cite{SemSte03} one can show
that only the terms $1$, $\exp(-4 r_i)$, $\exp(-2 r_i (1+\sigma))$,
$\exp(-2 r_i (2+2\sigma))$, and $\exp(-2 r_i (3+\sigma))$ actually appear,
but we do not need to use this fact.

As described in the Introduction,
we will adopt the
random effects approach and assume
that the $r_i$ are, in fact,
realizations of a sequence of independent,
identically distributed random variables
that we will denote by $(A_i)$.  

We are interested in finding such a sequence
$(A_i)$ and another independent, identically
distributed sequence $(B_i)$ such that
the distribution for the data induced by the
random choice of scale factors $(A_i)$ is
the same as that induced by the $(B_i)$
for {\bf another} choice of edge-length $\tau \ne \sigma$.

We thus have to find
positive random variables $A$ and $B$ 
with distinct distributions and distinct positive
constants $\sigma$ and $\tau$ with the property that
\[
\begin{split}
\bE[\exp(-2 A)] & = \bE[\exp(-2 B)]\\
\bE[\exp(-4 A)] & = \bE[\exp(-4 B)]\\
\bE[\exp(-2 \sigma A)] & = \bE[\exp(-2 \tau B)]\\
%\bE[\exp(-2 (1+\sigma)A)] & = \bE[\exp(-2 (1+\tau)B)]\\
%\bE[\exp(-2 (2+\sigma)A)] & = \bE[\exp(-2 (2+\tau)B)]\\
%\bE[\exp(-2 (3+\sigma)A)] & = \bE[\exp(-2 (3+\tau)B)]\\
%\bE[\exp(-2 (1+2 \sigma)A)] & = \bE[\exp(-2 (1+2 \tau)B)]\\
%\bE[\exp(-2 (2+2 \sigma)A)] & = \bE[\exp(-2 (2+2 \tau)B)]\\
& \cdots \\
\bE[\exp(-2 (3+2 \sigma)A)] & = \bE[\exp(-2 (3+2 \tau)B)].\\
\end{split}
\]

Take $A$ to have the distribution which has Laplace transform
\[
\bE[\exp(- \zeta A)]
=
\left\{\prod_{i=1}^9 (1+d_i \zeta)^{-1}\right\} 
(1+ h \zeta)^{-1} (1+ k \zeta)^{-1}
\]
for positive parameters $d_1,\ldots,d_7$, $h$, $k$.
Thus $A$ has the distribution of the sum of $9$ independent
exponential random variables with respective means
$d_1,\ldots,d_7$, $h$, $k$.  Take 
$B$ to have the distribution which has Laplace transform
\[
\bE[\exp(- \zeta B)]
=
\left\{\prod_{i=1}^7 (1+g_i \zeta)^{-1}\right\} 
(1+\ell \zeta)^{-2}
\]
for positive parameters $g_1,\ldots,g_7$, $\ell$.
Thus $B$ has the distribution of the sum of $9$
independent exponential random variables with respective means
$g_1,\ldots,g_7$, $\ell$, $\ell$.

Define  maps $P : \bR_+^{10} \rightarrow \bR_+^9$
and $Q : \bR_+^9 \rightarrow \bR_+^9$ by
\[
\begin{split}
P_1(\sigma, d_1, \ldots, d_7, h, k)
& = 
\left\{\prod_{i=1}^7 (1+2 d_i)\right\} 
(1+2 h) (1+2 k)\\
P_2(\sigma, d_1, \ldots, d_7, h, k)
& = 
\left\{\prod_{i=1}^7 (1+4 d_i)\right\} 
(1+4 h) (1+4 k)\\
P_3(\sigma, d_1, \ldots, d_7, h, k)
& = 
\left\{\prod_{i=1}^7 (1+2 d_i \sigma)\right\} 
(1+2 h \sigma) (1+2 k \sigma) \\
& \dots \\
P_9(\sigma, d_1, \ldots, d_7, h, k)
& = 
\left\{\prod_{i=1}^7 (1+2 d_i (3+2\sigma))\right\} 
(1+2 h (3+2 \sigma)) (1+2 k (3+2\sigma))\\
\end{split}
\]
and
\[
\begin{split}
Q_1(\tau, g_1, \ldots, g_7, \ell)
& = 
\left\{\prod_{i=1}^7 (1+2 g_i)\right\} 
(1+2 \ell)^2\\
Q_2(\tau, g_1, \ldots, g_7, \ell)
& = 
\left\{\prod_{i=1}^7 (1+4 g_i)\right\} 
(1+4 \ell)^2\\
Q_3(\tau, g_1, \ldots, g_7, \ell)
& = 
\left\{\prod_{i=1}^7 (1+2 g_i \tau)\right\} 
(1+2 \ell \tau)^2 \\
& \dots \\
Q_9(\tau, g_1, \ldots, g_7, \ell)
& = 
\left\{\prod_{i=1}^7 (1+2 g_i (3+2\tau))\right\} 
(1+2 \ell (3+2 \tau))^2.\\
\end{split}
\]
We want to show that
$
P(\sigma, d_1, \ldots, d_7, h, k) 
= 
Q(\tau, g_1, \ldots, g_7, \ell)
$
for some choice of parameters with $\sigma \ne \tau$.

Write $J(\sigma, d_1, \ldots, d_7, h, k)$ 
for the Jacobian matrix of the mapping
$
(d_1, \ldots, d_7, h, k) 
\mapsto 
P(\sigma, d_1, \ldots, d_7, h, k)
$
(thus, $J$ is a $9 \times 9$ matrix).  Write
$K(\tau, g_1, \ldots, g_7, \ell)$ for the Jacobian matrix of $Q$.
A straightforward check with a computer algebra package
such as {\em Mathematica} shows that the 
polynomials $\det J$ and $\det K$ are not identically 0.
(While the determinants could possibly be computed symbolically,
it is easier to compute the matrices symbolically, substitute
in appropriate integer values for the parameters, and use
exact integer arithmetic to compute the determinant for
those values:  For example,  
$\det J(2, 3, 4, 5, 6, 7, 8, 9, 10, 11) \ne 0$ and
$\det K(2, 3, 4, 5, 6, 7, 8, 9, 10) \ne 0$.)
Because these determinants are polynomials, the set of values
where $J$ (resp. $K$) is non-singular is a relatively
open subset of $\bR_+^{10}$ (resp. $\bR_+^9$)
with a closure that is all of $\bR_+^{10}$ (resp. $\bR_+^9$)
(that is, they are everywhere dense).

We can therefore find a point  
$(\bar \tau, \bar g_1, \ldots, \bar g_7, \bar \ell) $
in the interior of $\bR_+^9$ such that
\begin{itemize}
\item[(i)]
the matrix $K(\bar \tau, \bar g_1, \ldots, \bar g_7, \bar \ell)$
is non-singular and
\item[(ii)]
in any open neighborhood of
$(\bar g_1, \ldots, \bar g_7, \bar \ell, \bar \ell) \in \bR_+^9$
there are points $(d_1, \ldots, d_7, h, k)$ such that the matrix
$J(\bar \tau, d_1, \ldots, d_7, h, k)$ is non-singular.
\end{itemize}

By assumption (i) and the implicit function theorem
(see, for example, \cite{MR2003f:26001}),
the range of $Q$ contains an open neighborhood of
$Q(\bar \tau, \bar g_1, \ldots, \bar g_7, \bar \ell)$.
Note that 
$
P(\bar \tau, \bar g_1, \ldots, \bar g_7, \bar \ell, \bar \ell)
=
Q(\bar \tau, \bar g_1, \ldots, \bar g_7, \bar \ell),
$
and so for all points $(\sigma, d_1, \ldots, d_7, h, k)$
in some open neighborhood of 
$(\bar \tau, \bar g_1, \ldots, \bar g_7, \bar \ell, \bar \ell)$
we can find $(\tau, g_1, \ldots, g_7, \ell)$ such that
$ 
P(\sigma, d_1, \ldots, d_7, h, k) 
= 
Q(\tau, g_1, \ldots, g_7, \ell).
$

We will be done if we can show that it is not always the case
that $\sigma = \tau$ for such a solution. To see this,
we will fix $\sigma = \bar \tau$ and let 
$(d_1, \ldots, d_7, h, k)$ vary.  By assumption (ii)
and the implicit function theorem, the image
of any open neighborhood of  
$(\bar g_1, \ldots, \bar g_7, \bar \ell, \bar \ell)$ by the map
$
(d_1, \ldots, d_7, h, k) 
\mapsto 
P(\bar \tau, d_1, \ldots, d_7, h, k)
$
has non-empty interior.  However, the range of the map
$
(g_1, \ldots, g_7, \ell)
\mapsto
Q(\bar \tau, g_1, \ldots, g_7, \ell)
$
is at most $8$-dimensional, and, in particular, has empty interior.
Therefore, there certainly exists 
$(d_1, \ldots, d_7, h, k)$ such that
$
P(\bar \tau, d_1, \ldots, d_7, h, k) 
=
Q(\tau, g_1, \ldots, g_7, \ell)
$
for some  $(\tau, g_1, \ldots, g_7, \ell)$ with $\tau \ne \bar \tau$.

\bigskip
\noindent
{\bf Remark.} Note that if we take 
$g_1 = \dots = g_7 = \ell$, then we  
have a gamma distribution with shape parameter $9$.
Also, we could still produce the  unidentifiability phenomenon 
witnessed above if we raised all the
Laplace transforms to the same power $c>0$.  
In that case, setting $g_1 = \dots = g_7 = \ell$ would give 
a gamma distribution with shape parameter $9c$.  
Since the unidentifiability occurs on a dense set
of parameters $(g_1, \ldots, g_7, \ell)$, any
gamma distribution 
will have distributions arbitrarily close to it
that exhibit the phenomenon.

\section{Conclusions and Future Research}
\label{section:conclusion}
The example we have given
shows that the attempt to achieve
identifiability and reasonable inference of
edge-lengths in the rates--across--sites model
by using random scale factors that come from some
common distribution can be problematic.

The gamma distributions `work', but distributions
arbitrarily close to any given gamma 
with smooth, unimodal densities and finite moments of all orders don't.
Using the gamma family is thus not just a matter of working
with distributions that have enough flexibility to
capture reasonable variation in rates.   Rather,
identifiability of edge-lengths for the gamma family
relies on quite specific features of members of that family
that are not shared by equally reasonable distributions.

The result has consequences for the estimation of times
at internal nodes, since if edge-lengths cannot be estimated,
then neither can the dates (since the edge-length is a product
of the elapsed time on the edge, and the equilibrium expected rate of evolution
for on that edge). 
%It is also reasonable to consider the question of whether
%model trees are identifiable under generic cases of these
%simple distributions of rates across sites.  Our result does not answer
%this, and so further research is necessary.

Finally, although our main result is theoretical, its consequences 
can be tested in simulation.  To date, few (if any)
such studies have been done that have not presumed that the
rates are distributed by a gamma distribution, or a distribution
consisting of some invariable sites, and the remaining sites
evolving under a gamma distribution.  This also reflects the 
 implicit  belief that the assumption of
a gamma distribution is acceptable. We hope this paper will help
encourage researchers to reconsider this assumption.

\bigskip\noindent
{\bf Acknowledgments}  The authors thank the anonymous referees
for a number of very helpful suggestions and observations.
TW acknowledges the support of the
Program for Evolutionary Dynamics at Harvard, and the
Institute for Cellular and Molecular Biology at UT-Austin.

\providecommand{\bysame}{\leavevmode\hbox to3em{\hrulefill}\thinspace}
\providecommand{\MR}{\relax\ifhmode\unskip\space\fi MR }
% \MRhref is called by the amsart/book/proc definition of \MR.
\providecommand{\MRhref}[2]{%
  \href{http://www.ams.org/mathscinet-getitem?mr=#1}{#2}
}
\providecommand{\href}[2]{#2}


\begin{thebibliography}{OMHO94}

\bibitem[Cha96]{Cha96}
J.~Chang, \emph{Full reconstruction of markov models on evolutionary trees:
  identifiability and consistency}, Mathematical Biosciences \textbf{137}
  (1996), 51--73.

\bibitem[Fel78]{Fel78}
J.~Felsenstein, \emph{Cases in which parsimony and compatibility methods will
  be positively misleading}, Syst. Zool. \textbf{27} (1978), 401--410.

\bibitem[Fel04]{Fel04}
\bysame, \emph{Inferring phylogenies}, Sinauer Associates, Sunderland,
  Massachusetts, 2004.

\bibitem[GG03]{GuindonOlivier2003}
S.~Guindon and O.~Gascuel, \emph{A simple, fast, and accurate algorithm to
  estimate large phylogenies by maximum likelihood}, Syst. Biol. \textbf{52}
  (2003), no.~5, 696--704.

\bibitem[GS01]{GriSti01}
G.R. Grimmett and D.R. Stirzaker, \emph{Probability and random processes},
  third ed., Oxford University Press, New York, 2001.

\bibitem[HKY87]{HasKisYan87}
M.~Hasegawa, H.~Kishino, and T.~Yano, \emph{Man's place in {H}omonoidea as
  inferred from molecular clocks of {DNA}}, J. Mol. Evol. \textbf{26} (1987),
  132--147.

\bibitem[HR01]{HuelsenbeckRonquist2001}
J.P. Huelsenbeck and R.~Ronquist, \emph{{MrBayes: B}ayesian inference of
  phylogeny.}, Bioinformatics \textbf{17} (2001), 754--755.

\bibitem[JN90]{JinNei90}
L.~Jin and M.~Nei, \emph{Limitations of evolutionary parsimony methods of
  phylogenetic analysis}, Mol. Biol. Evol. \textbf{7} (1990), 82--102.

\bibitem[KP02]{MR2003f:26001}
S.G. Krantz and H.R. Parks, \emph{The implicit function theorem}, Birkh\"auser
  Boston Inc., Boston, MA, 2002, History, theory, and applications.
  \MR{2003f:26001}

\bibitem[Lew98]{Lewis98}
P.~Lewis, \emph{A genetic algorithm for maximum likelihood phylogeny inference
  using nucleotide sequence data}, Mol. Biol. Evol. \textbf{15} (1998),
  277--283.

\bibitem[NCF76]{NeiChaFue76}
M.~Nei, R.~Chakraborty, and P.A. Fuerst, \emph{Infinite allele model with
  varying mutation rate}, Proc. Natl. Acad. Sci. USA \textbf{73} (1976),
  4164--4168.

\bibitem[Ols87]{Ols87}
G.J. Olsen, \emph{Earliest phylogenetic branchings: {C}omparing {rRNA}-based
  evolutionary trees inferred with various techniques}, Cold Spring Harbor
  Symposia on Quantitative Biology \textbf{52} (1987), 825--837.

\bibitem[OMHO94]{OlsenMHO94}
G.~Olsen, H.~Matsuda, R.~Hagstrom, and R.~Overbeek, \emph{{Fast{DNA}ml: {A}
  tool for construction of phylogenetic trees of {DNA} sequences using maximum
  likelihood}}, Computations in Applied Biosciences \textbf{10} (1994), no.~1,
  41--48.

\bibitem[PM00]{KosakovskyPondMuse2000}
S.~L.~Kosakovsky Pond and S.~Muse, \emph{Hyphy package distribution and
  documentation page}, 2000, Distributed by the authors at {\tt
  http://www.hyphy.org}.

\bibitem[Ree92]{Ree92}
J.H. Reeves, \emph{Heterogeneity in the substitution process of amino acid
  sites of proteins coded for by mitochondrial {DNA}}, J. Mol. Evol. (1992),
  17--31.

\bibitem[Rog01]{Rog01}
J.S. Rogers, \emph{Maximum likelihood estimation of phylogenetic trees is
  consistent when substitution rates vary according to the invariable sites
  plus gamma distribution}, Syst. Biol. \textbf{50} (2001), 713--722.

\bibitem[SOWH96]{SwoOlsWadHil96}
D.L. Swofford, G.J. Olsen, P.J. Waddell, and D.M. Hillis, \emph{Phylogenetic
  inference}, Molecular Systematics (D.M. Hillis, C.~Moritz, and B.K. Mable,
  eds.), Sinauer Associates, Sunderland, Massachusetts, 1996.

\bibitem[SS03]{SemSte03}
C.~Semple and M.~Steel, \emph{Phylogenetics}, Oxford Lecture Series in
  Mathematics and its Applications, vol.~24, Oxford University Press, Oxford,
  2003.

\bibitem[SSH94]{SteSzeHen94}
M.A. Steel, L.A. Sz\'ekely, and M.D. Hendy, \emph{Reconstructing trees when
  sequence sites evolve at variable rates}, J. Comp. Biol. \textbf{1} (1994),
  153--163.

\bibitem[Ste94]{Ste94}
M.A. Steel, \emph{Recovering a tree from the leaf colourations it generates
  under a markov model}, Applied Mathematics Letters \textbf{7} (1994), 19--24.

\bibitem[Swo96]{Swofford96}
D.~Swofford, \emph{{PAUP*:} phylogenetic analysis using parsimony (and other
  methods), version 4.0}.

\bibitem[TS97]{tuffley.steel}
C.~Tuffley and M.~Steel, \emph{Links between maximum likelihood and maximum
  parsimony under a simple model of site substitution}, Bulletin of
  Mathematical Biology \textbf{59} (1997), 581--607.

\bibitem[UC71]{UzzCor71}
T.~Uzzell and K.W. Corbin, \emph{Fitting discrete probability distributions to
  evolutionary events}, Science \textbf{72} (1971), 1089--1096.

\bibitem[Yan96]{Yan93}
Z.~Yang, \emph{Maximum-likelihood estimation of phylogeny from {DNA} sequences
  when substitution rates differ over sites}, Mol. Biol. Evol. \textbf{10}
  (1996), 1396--1401.

\end{thebibliography}
\end{document}